\input{aipcheck}

\documentclass[
  ,final            % use final for the camera ready runs
 ]
{aipproc}

\layoutstyle{8x11single}

\usepackage{bm}

\begin{document}

\title{Rheology of Dense Sheared Granular Liquids}

\classification{45.70.-n, 61.20.Lc, 64.70.ps, 83.50.Ax, 83.60.Fg}
%<Replace this text with PACS numbers; choose from this list:
%\texttt{http://www.aip..org/pacs/index.html}>
\keywords      {Granular matter, Liquid theory, Mode coupling theory}

\author{Koshiro Suzuki}{
  address={Canon Inc., 30-2 Shimomaruko 3-chome, Ohta-ku, Tokyo
  146-8501, Japan}
}

\author{Hisao Hayakawa}{
  address={Yukawa Institute for Theoretical Physics, Kyoto University,
  Kitashirakawa Oiwake-cho, Kyoto 606-8502, Japan}
}

\begin{abstract}
The rheology of dense sheared granular liquids is investigated based on
the mode-coupling theory (MCT).
This extended MCT includes correlations for the density-current mode as
well as the density-density correlation mode, and a self-consistent
coupling equation for the energy balance condition.
The extended MCT exhibits disappearance of the two-step relaxation of
the density-density correlation function, and also successfully
reproduces the density dependence of the shear viscosity for volume
fractions between 0.50 and 0.60, if we shift the density.
However, it predicts unphysical tendency for the granular temperature.
The cause of this drawback and the possibilities of its amendment are
discussed.
\end{abstract}

\maketitle

%%%%%%%%%%%%%%%%%%%%%%%%%%%%%%%%%%%%%%%%%%%%
%% MAINMATTER
%%%%%%%%%%%%%%%%%%%%%%%%%%%%%%%%%%%%%%%%%%%%

\section{Introduction}

Establishing a macroscopic description of granular materials has been a
long-term challenge for both science and engineering.
The problem extends to a vast range, from creep motion or force chain
dynamics of frictional particles, two-phase flow of air-fluidized beds,
to nonequilibrium transport of sheared granular flows \cite{JNB1996}.
Similar to solid-liquid transitions, critical features of the jamming
transition in the vicinity of the random close packing and their
relation to the glass transition have attracted much interest in the
last decade \cite{LN1998, IBS2012}.
Even when we focus only on the classical problem of the flow properties
well below the jamming transition density $\varphi_{\mathrm{J}}$, which
can be traced back to Bagnold's work \cite{Bagnold1954}, we have not yet
understood the rheological properties of dense granular flows.
One of the remarkable achievements is the extension of the
Boltzmann-Enskog kinetic theory to inelastic hard disks and spheres
\cite{JR1985-2}, which stimulated the following works \cite{GD1999,
SH2007}.
%
%Up to present, the kinetic theory is the unique continuum description of
%granular flows derived from first principles \cite{BP}.
%
However, it has been recognized that the kinetic theory breaks down at
high densities with volume fraction $\varphi > 0.5$ \cite{MN2007,
JB2010}, since there exists correlated motions of grains.
Thus, a liquid theory which contains the effect of granular correlations
is expected to be constructed for the regime $0.5 < \varphi <
\varphi_{\mathrm{J}}$.

On the other hand, a continuum description with long-time
correlations has been constructed for thermal glassy liquids.
The mode-coupling theory (MCT) exhibits the two-step relaxation of
density correlation functions characteristic of glasses \cite{Goetze}.
Although MCT has presented remarkable success, it is marred with
problems;
for instance, it predicts a non-ergodic transition which is not observed
in experiments \cite{KA1994}, or its critical density
$\varphi_{\mathrm{MCT}}=0.516$ \cite{BGS1984} is far below the glass
transition density, which is observed to be $\varphi_{g} \simeq
0.58-0.60$ in numerical simulations \cite{GGF1976}.

Despite these problems, it is still tempting to extend MCT to granular
materials, because it might be the simplest method to include
correlations.
In MCT, these correlations are encoded in a memory kernel, which is
absent in the kinetic theory.
The extension of MCT to randomly driven granular systems has been
proposed and analyzed \cite{KSZ2013}.
However, the physics of sheared flows is completely distinct from random
driving cases, and its study has to be addressed independently.
Indeed, the characteristic plateau in the density correlation function
for both glassy and randomly driven granular systems does not exist in
sheared granular liquids \cite{CC2009}.
The extension of MCT to sheared granular liquids can be traced back to
Ref.~\cite{HO2008}, and an explicit calculation of time correlation
functions has been reported in Ref.~\cite{SH2013-3} without considering
the energy balance condition.
However, the problem of rheology and the characterization of the steady
state have not been addressed so far.
In this article, we demonstrate that the extended MCT reproduces the
results of molecular dynamics (MD) simulations for the relaxation of
time correlation functions and the density dependence of the shear
viscosity.

The organization of this paper is as follows. 
We first explain the microscopic set up of the theory, in particular the
Liouville equation and the steady-state condition.
Then we derive the mode-coupling equations, without specifying the
initial distribution function.
We perform a concrete calculation for the case of a canonical initial
distribution and evaluate the validity of the results.
Finally, we discuss the problem of the formulation and the possibilities
of its amendment.

\vspace{-1em}
\section{Microscopic setup}
\subsection{Equations of Motion}
The starting point of our theory is an assembly of $N$ identical soft,
smooth, inelastic spheres with mass $m$ and diameter $d$, contained in a
box of volume $V$.
Thus, the number density is given by $n=N/V$.
%
%
%\footnote{To incorporate the Coulomb friction is a difficult issue, due
%to the singularity of the friction at the stationary state.}.
%
The interaction between the spheres are elastic as well as inelastic
(dissipative), both of which are repulsive and emerge at contact.
The system is subjected to a bulk shearing, which is uniform and
constant with shear rate $\dot\gamma$
%
%\footnote{We should analyze a shear driven by a physical boundary, but
%the sheared region is localized near the boundary. 
%%
%As for the problem of time-dependent shear, we may extend the
%formulation for colloidal suspensions reported in Ref.~\cite{BCF2012},
%which is one of our future tasks.}.
%
.
At first the system is equilibrated without shear and dissipation.
Then, at $t=t_{0}$ $(<0)$, both of them are switched on, which
eventually leads the system to a nonequilibrium steady state.
The Newtonian equation of motion for the $i$th sphere ($i=1,\cdots,N$)
is given by the following set of Sllod equations \cite{EM}:
\vspace{-1.0em}
\begin{eqnarray}
\dot{\bm{r}}_{i}(t)
&=&
\frac{\bm{p}_{i}(t)}{m} 
+ 
\dot{\bm{\gamma}}\cdot\bm{r}_{i}(t),
\label{eq:r}
\\
\dot{\bm{p}}_{i}(t)
&=&
\bm{F}^{(\mathrm{el})}_{i}(t) 
+
\bm{F}^{(\mathrm{vis})}_{i}(t) 
-
\dot{\bm{\gamma}}\cdot\bm{p}_{i}(t).
\label{eq:p}
\end{eqnarray}
Here, $\bm{\Gamma}(t) \equiv \{ \bm{r}_{i}(t), \bm{p}_{i}(t)
\}_{i=1}^{N}$ is a set of the positions and the momenta of all the
grains at time $t$, $\dot{\bm{\gamma}}$ is the shear rate tensor whose
components are assumed to be given by $\dot{\gamma}_{\mu\nu}=\dot\gamma
\delta_{\mu x} \delta_{\nu y}$, and $\bm{F}_{i}^{(\mathrm{el})}(t)$,
$\bm{F}_{i}^{(\mathrm{vis})}(t)$ are the elastic and dissipative
interactions, respectively.
The Greek indices $\mu, \nu, \lambda, \cdots$ denote spatial components
$\{ x, y, z \}$, and the rule for the summation over repeated indices is
implied.
Here, the elastic force is given by $\bm{F}_{i}^{(\mathrm{el})}(t)
\equiv - \sum_{j\neq i} \Theta(d-r_{ij}) \partial u(r_{ij}(t)) /
\partial \bm{r}_{ij}(t)$, where $\bm{r}_{ij}(t) \equiv \bm{r}_{i}(t) -
\bm{r}_{j}(t)$ and $r_{ij}(t) \equiv |\bm{r}_{ij}(t)|$ are,
respectively, the relative position and the distance between the $i$th
and $j$th spheres, $u(r)$ is the two-body potential, and $\Theta(x)$ is
a step function which is 1 for $x>0$ and 0 for other cases.
Although a realistic potential might be Hertzian, it is assumed to be
harmonic for simplicity.
The dissipative force is given by $\bm{F}_{i}^{(\mathrm{vis})}(t) \equiv
- \zeta \sum_{j\neq i} \Theta(d-r_{ij}) \hat{\bm{r}}_{ij}
(\dot{\bm{r}}_{ij} \cdot \hat{\bm{r}}_{ij})$, where $\hat{\bm{r}} \equiv
\bm{r}/|\bm{r}|$ is a unit vector and $\zeta$ is a viscous constant
corresponding to the harmonic potential.

\vspace{-0.5em}
\subsection{Liouville Equation}
To formulate a theory, we rewrite the equations of motion,
Eqs.~(\ref{eq:r}) and (\ref{eq:p}), to the form of the Liouville
equation \cite{EM}.
In this formulation, the equation of motion for an arbitrary phase-space
variable $A(\bm{\Gamma})$ casts into the form
\vspace{-0.5em}
\begin{eqnarray}
\frac{d}{dt}
A(\bm{\Gamma}(t)) 
=
i\mathcal{L}(\bm{\Gamma})
A(\bm{\Gamma}(t)),
\label{eq:Liouv_A}
\end{eqnarray}
where the operator $i\mathcal{L}(\bm{\Gamma})$ is referred to as the
Liouvillian.
The explicit form of $i\mathcal{L}(\bm{\Gamma})$ for Eqs.~(\ref{eq:r})
and (\ref{eq:p}) is given by
\begin{equation}
i\mathcal{L}(\bm{\Gamma})
=
i\mathcal{L}^{(\mathrm{el})}(\bm{\Gamma})
+
i\mathcal{L}_{\dot\gamma}(\bm{\Gamma})
+
i\mathcal{L}^{(\mathrm{vis})}(\bm{\Gamma}),
\end{equation}
where the elastic part $i\mathcal{L}^{(\mathrm{el})}(\bm{\Gamma})$, the
shear part $i\mathcal{L}_{\dot\gamma}(\bm{\Gamma})$, and the viscous
part $i\mathcal{L}^{(\mathrm{vis})}(\bm{\Gamma})$ are given by
%
%\begin{eqnarray}
$i\mathcal{L}^{(\mathrm{el})}(\bm{\Gamma})
=
\sum_{i=1}^{N}
\left[
\frac{\bm{p}_{i}}{m} \cdot \frac{\partial}{\partial \bm{r}_{i}}
+
\bm{F}_{i}^{(\mathrm{el})} \cdot \frac{\partial}{\partial \bm{p}_{i}}
\right],
$
$i\mathcal{L}_{\dot\gamma}(\bm{\Gamma})
=
\sum_{i=1}^{N}
\left[
\left( \bm{\dot\gamma}\cdot\bm{r}_{i} \right) \cdot \frac{\partial}{\partial \bm{r}_{i}}
-
\left( \bm{\dot\gamma}\cdot\bm{p}_{i} \right) \cdot \frac{\partial}{\partial \bm{p}_{i}}
\right],
$
$i\mathcal{L}^{(\mathrm{vis})}(\bm{\Gamma})
=
\sum_{i=1}^{N}
\bm{F}_{i}^{(\mathrm{vis})} \cdot \frac{\partial}{\partial \bm{p}_{i}}.
$
On the other hand, the equation for the phase-space distribution
function $\rho(\bm{\Gamma},t)$ reads
\vspace{-0.5em}
\begin{eqnarray}
\frac{\partial}{\partial t}
\rho(\bm{\Gamma},t)
=
-i\mathcal{L}^{\dagger}(\bm{\Gamma}), 
\label{eq:Liouv_rho}
\end{eqnarray}
where the adjoint Liouvillian $i\mathcal{L}^{\dagger}(\bm{\Gamma})$ is
defined as
$%\begin{eqnarray}
i\mathcal{L}^{\dagger}(\bm{\Gamma})
=
i\mathcal{L}(\bm{\Gamma})
+
\Lambda(\bm{\Gamma}),
$ %\end{eqnarray}
together with the phase-space contraction factor,
$%\begin{eqnarray}
\Lambda(\bm{\Gamma}) 
=
%\frac{\partial}{\partial \bm{\Gamma}}
\partial/\partial \bm{\Gamma}
\cdot
\dot{\bm{\Gamma}}
=
-%\frac{\zeta}{m} 
(\zeta/m)
\sum_{\langle i,j \rangle}
\Theta(d-r_{ij}) < 0.
$ %\end{eqnarray}
Here, we denote $\sum_{i=1}^{N}\sum_{j\neq i}\cdots$ as $\sum_{\langle
i,j \rangle} \cdots$.
The formal solution of Eqs.~(\ref{eq:Liouv_A}) and (\ref{eq:Liouv_rho})
are given by
\vspace{-0.5em}
\begin{eqnarray}
A(\bm{\Gamma}(t)) 
=
e^{i\mathcal{L}t}
A(\bm{\Gamma}(0)),
\hspace{1em}
\rho(\bm{\Gamma},t)
=
e^{-i\mathcal{L}^{\dagger}t}
\rho(\bm{\Gamma},0).
\label{eq:rho_formal}
\end{eqnarray}
To proceed, it is convenient to rewrite $\rho(\bm{\Gamma},t)$ in
Eq.~(\ref{eq:rho_formal}), by use of an identity \cite{EM}, as
\begin{eqnarray}
\rho(\bm{\Gamma},t)
=
\rho_{\mathrm{ini}}(\bm{\Gamma})
+
\int_{0}^{t} ds \,
e^{-i\mathcal{L}^{\dagger}s}
\left( -i\mathcal{L}^{\dagger}\right)
\rho_{\mathrm{ini}}(\bm{\Gamma}),
\end{eqnarray}
where $\rho_{\mathrm{ini}}(\bm{\Gamma})\equiv \rho(\bm{\Gamma},0)$ is
the initial distribution function.
Then, we can introduce the nonequilibrium work function,
$\Omega(\bm{\Gamma})$, as the eigenvalue of $i\mathcal{L}^{\dagger}$:
\vspace{-0.5em}
\begin{eqnarray}
i\mathcal{L}^{\dagger}
\rho_{\mathrm{ini}}(\bm{\Gamma})
=
-
\rho_{\mathrm{ini}}(\bm{\Gamma})
\Omega(\bm{\Gamma}), 
\label{eq:Omega_def}
\end{eqnarray}
from which we obtain
\vspace{-0.5em}
\begin{eqnarray}
\rho(\bm{\Gamma},t)
=
\rho_{\mathrm{ini}}(\bm{\Gamma})
+
\int_{0}^{t} ds \,
e^{-i\mathcal{L}^{\dagger}s}
\left[
\rho_{\mathrm{ini}}(\bm{\Gamma})
\Omega(\bm{\Gamma})
\right].
\label{eq:rho_dyson}
\end{eqnarray}
Note that $\Omega(\bm{\Gamma})$ includes the nonequilibrium information
of the dynamics of the system, which is caused by shearing and
dissipation.
%
%Note also that Eq.~(\ref{eq:rho_dyson}) can be identically rewritten as
%\cite{EM}
%%
%\begin{eqnarray}
%\rho(\bm{\Gamma},t) 
%=
%\exp \left[
%\int_{0}^{t} d\tau \,
%\Omega(-\tau)
%\right]
%\rho_{\mathrm{ini}}(\bm{\Gamma}).
%\label{eq:rho_kawasaki}
%\end{eqnarray}
%%
In the remainder, we denote the ensemble average with respect to
$\rho_{\mathrm{ini}}(\bm{\Gamma})$ by
\begin{eqnarray}
\left\langle \cdots \right\rangle
=
\int d\bm{\Gamma} \,
\rho_{\mathrm{ini}}(\bm{\Gamma}) \cdots.
\label{eq:ave_def}
\end{eqnarray}

The nonequilibrium ensemble average of a phase-space variable
$A(\bm{\Gamma}(t))$, which we denote $\langle A(t) \rangle$, can be
expressed as
\begin{equation}
\left\langle
A(t)
\right\rangle
=
\int d\bm{\Gamma} \,
\rho_{\mathrm{ini}}(\bm{\Gamma})
A(\bm{\Gamma}(t))
=
\int d\bm{\Gamma} \,
\rho(\bm\Gamma,t)
A(\bm{\Gamma}(0)).
\label{eq:At}
\end{equation}
The equivalence of the two expressions in Eq.~(\ref{eq:At}) can be shown
by the adjoint relation of the Liouvillians \cite{EM}.
From Eqs.~(\ref{eq:rho_dyson}), (\ref{eq:At}), and the adjoint relation,
we obtain
$%\begin{eqnarray}
\left\langle
A(t)
\right\rangle
=
\left\langle
A(0)
\right\rangle
+
\int_{0}^{t} ds \,
\left\langle
A(s)
\Omega(0)
\right\rangle,
$ %\end{eqnarray}
which is referred to as the generalized Green-Kubo formula \cite{EM}.
The steady-state average, $A_{\infty}$, is obtained from this formula as
\begin{eqnarray}
A_{\infty}
=
\left\langle
A(0)
\right\rangle
+
\int_{0}^{\infty} ds \,
\left\langle
A(s)
\Omega(0)
\right\rangle.
\label{eq:ss}
\end{eqnarray}

\vspace{-1.0em}
\subsection{Steady-State Condition}
The steady state of a sheared granular flow is characterized by the
balance of energy.
The internal energy $H_{0}(\bm{\Gamma})$ is given 
$%\begin{eqnarray}
H_{0}(\bm{\Gamma}) 
=
\sum_{i=1}^{N}
\left[
%\frac{\bm{p}_{i}^2}{2m}
\bm{p}_{i}^2/2m
+
\sum_{j\neq i}
u(r_{ij})
\right].
$ %\end{eqnarray}
The following equality can be shown from Eqs.~(\ref{eq:r}) and
(\ref{eq:p}),
\begin{eqnarray}
\dot{H}_{0}(\bm{\Gamma})
=
-\dot\gamma
\sigma_{xy}(\bm{\Gamma})
-
2\mathcal{R}(\bm{\Gamma}),
\label{eq:dH0}
\end{eqnarray}
where
$%\begin{eqnarray}
\sigma_{xy}(\bm{\Gamma}) 
=
\frac{1}{V}
\sum_{i=1}^{N}
\left[
%\frac{p_{i}^{\mu}p_{i}^{\nu}}{m}
p_{i}^{x}p_{i}^{y}/m
+
y_{i}F_{i}^{x}
\right]
$ %\end{eqnarray}
is the shear stress and 
$%\begin{equation}
\mathcal{R}(\bm{\Gamma}) 
=
-\frac{1}{4}
\sum_{\langle i,j \rangle}
\dot{\bm{r}}_{ij} \cdot \bm{F}_{ij}^{(\mathrm{vis})}
=
\frac{\zeta}{4}
\sum_{\langle i,j \rangle}
\Theta(d-r_{ij})
\left( \dot{\bm{r}}_{ij} \cdot \hat{\bm{r}}_{ij}\right)^2
$ %\end{equation}
is Rayleigh's dissipation function.
From Eq.~(\ref{eq:dH0}) the steady state is characterized by the
condition,
%
%\vspace{-0.5em}
\begin{eqnarray}
\dot\gamma 
\left( \sigma_{xy} \right)_{\infty}
+
2\mathcal{R}_{\infty}
=
0, 
\label{eq:ss2}
\end{eqnarray}
where the steady-state values $(\sigma_{xy})_{\infty}$ and
$\mathcal{R}_{\infty}$ can be evaluated by Eq.~(\ref{eq:ss}).
Equation~(\ref{eq:ss2}) implies the balance between the heating due to
shear and the cooling due to dissipation.
As will be discussed later, the time evolution of the temperature is not
included in Eq.~(\ref{eq:ss2}) in our framework, and hence it will be
imposed as a condition which determines the steady-state temperature.

\vspace{-1.0em}
\section{Sheared granular MCT}
So far the formulation is exact but not solvable.
To obtain a calculable theory for the low-frequency modes, Mori
equations are approximated by MCT to derive a closed set of equations.
MCT is also applied to the generalized Green-Kubo formula to calculate
the shear stress and the energy dissipation rate.
First we briefly review the essence of this procedure \cite{SH2013-3}.

\vspace{-0.5em}
\subsection{MCT Equations}

We identify the density and the current density fluctuations, which are
$n_{\bm{q}} \equiv \sum_{i=1}^{N}e^{i\bm{q}\cdot\bm{r}_{i}} -
N\delta_{\bm{q},0}$ and $j_{\bm{q}}^{\lambda} \equiv \sum_{i=1}^{N}
\left( p_{i}^{\lambda}/m \right) e^{i\bm{q}\cdot\bm{r}_{i}}$ in Fourier
space, as the slow modes.
In standard granular hydrodynamics, the temperature fluctuation should
also be considered on the same grounds as $n_{\bm{q}}$ and
$j_{\bm{q}}^{\lambda}$.
However, we attempt to formulate a theory without the temperature
fluctuation, since otherwise it will be too complicated and almost
untractable. 
How to compensate the dynamics of the temperature will be discussed
below.

The Mori equations for $n_{\bm{q}}$ and $j_{\bm{q}}^{\lambda}$ are
obtained by applying the projection operator 
\begin{eqnarray}
\mathcal{P}(t)X 
=
\sum_{\bm{k}} 
\frac{\langle Xn_{\bm{k}(t)}^{*} \rangle}{NS_{k(t)}} 
n_{\bm{k}(t)} 
+ 
\sum_{\bm{k}} 
\frac{\langle X j_{\bm{k}(t)}^{\mu\,*} \rangle}{Nv_{T}^2}
j_{\bm{k}(t)}^{\mu},
\end{eqnarray}
where $X(\bm{\Gamma})$ is an arbitrary
phase-space variable.
Here, $S_{k}=\langle n_{\bm{k}}n_{\bm{k}}^{*}\rangle /N$ is the static
structure factor, $v_T^2 = \langle
j_{\bm{k}}^{\lambda}j_{\bm{k}}^{\lambda\,*} \rangle/(3N)$ is the
equal-time correlation of the current, and $\bm{k}(t) \equiv (k_x, k_y -
\dot\gamma t k_x, k_z)$ is the wave vector in the sheared frame.
Note that $\langle \cdots \rangle$ denotes the ensemble average with
respect to the initial distribution function, Eq.~(\ref{eq:ave_def}).
The orthogonality of $n_{\bm{k}}$ and $j_{\bm{k}}^{\lambda}$,
i.e. $\langle n_{\bm{k}}j_{\bm{k}}^{\lambda *}\rangle =0$, is necessary
for the idempotency, $\mathcal{P}(t)^2=\mathcal{P}(t)$.
We assume that $\rho_{\mathrm{ini}}(\bm{\Gamma})$ is even with respect
to the momentum, which is sufficient for $\langle
n_{\bm{k}}j_{\bm{k}}^{\lambda *}\rangle =0$.
Then, to obtain a closure for these equations, the second projection
operator 
\vspace{-0.5em}
\begin{eqnarray}
\mathcal{P}_{\mathrm{mc}}(t) = \mathcal{P}_{nn}(t) +
\mathcal{P}_{nj}(t),
\label{eq:Pmc}
\end{eqnarray}
where
\vspace{-0.5em}
\begin{eqnarray}
\mathcal{P}_{nn}(t) 
=
\sum_{\bm{k}>\bm{p}} 
\frac{\langle Xn_{\bm{k}(t)}^{*}n_{\bm{p}(t)}^{*}\rangle}{N^2
S_{k(t)}S_{p(t)}}
n_{\bm{k}(t)}n_{\bm{p}(t)},
\hspace{1em}
\mathcal{P}_{nj}(t) 
=
\sum_{\bm{k},\bm{p}}
\frac{\langle Xn_{\bm{k}(t)}^{*}j_{\bm{p}(t)}^{\mu\,*}\rangle}{N^2
S_{k(t)} v_{T}^{2}}
n_{\bm{k}(t)}j_{\bm{p}(t)}^{\mu},
\label{eq:PnnPnj}
\end{eqnarray}
is applied to the memory kernels, together with the factorization
approximation.
Then we obtain a set of coupled equations for four time correlation
functions, $\Phi_{\bm{q}}(t) \equiv \left\langle n_{\bm{q}(t)}(t)
n_{\bm{q}}(0)^{*} \right\rangle /N$, $H_{\bm{q}}^{\lambda}(t) \equiv i
\left\langle j_{\bm{q}(t)}^{\lambda}(t) n_{\bm{q}}(0)^{*} \right\rangle
/N$, $\bar{H}_{\bm{q}}^{\lambda}(t) \equiv i \left\langle
n_{\bm{q}(t)}(t) j_{\bm{q}}^{\lambda}(0)^{*} \right\rangle /N$, and
$C_{\bm{q}}^{\mu\nu}(t) \equiv \left\langle j_{\bm{q}(t)}^{\mu}(t)
j_{\bm{q}}^{\nu}(0)^{*} \right\rangle /N$.
However, we might expect that anisotropic effects are subdominant,
considering that sheared colloidal systems are almost isotropic
\cite{MRY2004, HWCF2009}.
Hence, we resort to the {\it isotropic approximation}, and reduce the
time correlation functions to two scalar functions, $\Phi_{q}(t)$ and
$\Psi_{q}(t)$, which depend only on the modulus of the wave vector, $q
\equiv |\bm{q}|$.
Here, $\Psi_{q}(t)$ is introduced by the definition
$\bar{H}_{\bm{q}}^{\lambda}(t) \equiv - q^{\lambda}(t) \Psi_{q}(t)$, and
hence is essentially the density-current correlation.
These two functions are related to the remaining two by the relations
$H_{\bm{q}}^{\lambda}(t) \simeq [ q^{\lambda}(t) /q(t)^2] \frac{d}{dt}
\Phi_{q}(t)$ and $C_{\bm{q}}^{\mu\nu}(t) \simeq \delta^{\mu\nu} \left\{
\frac{d}{dt} \Psi_{q}(t) + \frac{1}{2} \left[ \Psi_{q}(t)/q(t)^2 \right]
\frac{d}{dt} q(t)^2 \right\}$. 
It should be noted that, while $H_{\bm{q}}^{\lambda}(t)$ is obtained by
the time derivative of $\Phi_{\bm{q}}(t)$, $\bar{H}_{\bm{q}}(t)$ cannot
be derived from $\Phi_{\bm{q}}(t)$;
i.e., the minimal set of time correlation functions is $\{\Phi_{q},
\Psi_{q}\}$.
Then, the MCT equations for $\Phi_{q}(t)$ and $\Psi_{q}(t)$ read
\begin{eqnarray}
\frac{d^2}{dt^2} \Phi_{q}(t)
&=&
-Z_{q(t)} \Phi_{q}(t)
-A_{q(t)}\frac{d}{dt}\Phi_{q}(t)
%\nonumber \\
%%
%&&
-\int_{0}^{t} ds \, M_{\bm{q}(s)}(t-s) \frac{d}{ds} \Phi_{q}(s),
\label{eq:Phi}
\\
\frac{d^2}{dt^2} \Psi_{q}(t)
&=&
-\frac{1}{3} Z_{q(t)} \Psi_{q}(t)
-\frac{1}{3} A_{q(t)}^{\lambda\lambda} \frac{d}{dt}\Psi_{q}(t)
%\nonumber \\
%%
%&&
-\frac{1}{3}\int_{0}^{t} ds \, M_{\bm{q}(s)}^{\lambda\lambda}(t-s)
\frac{d}{ds} \Psi_{q}(s),
\label{eq:Psi}
\end{eqnarray}
where the memory kernels $M_{\bm{q}}(t)$ and
$M_{\bm{q}}^{\lambda\lambda}(t)$ are quadratic forms in terms of the
time correlation functions.
The coefficients $Z_{q}$, $A_{q}$, $A_{q}^{\lambda\lambda}$ and the
coefficients of the quadratic forms in the memory kernels are equal-time
correlations, which originate from the projection operators.
Hence, their explicit forms depend on the choice of
$\rho_{\mathrm{ini}}(\bm{\Gamma})$, which we will discuss later.
We simply note here that the memory kernels consist of elastic and
dissipative terms.
For instance, $M_{\bm{q}}(t)$ has the following form,
$%\begin{eqnarray}
M_{\bm{q}}(t) 
= 
M_{\bm{q}}^{(\mathrm{el})}(t) +
%\frac{\zeta}{m \dot\gamma}
(\zeta/m\dot\gamma)
M_{\bm{q}}^{(\mathrm{vis})}(t),
$ %\end{eqnarray}
where $M_{\bm{q}}^{(\mathrm{el})}(t)$ is the conventional term which
induces the glass transition for sheared non-dissipative particles
\cite{FC2009}, and $M_{\bm{q}}^{(\mathrm{vis})}(t)$ represents the
dissipative terms.
These features also hold for the kernel
$M_{\bm{q}}^{\lambda\lambda}(t)$.
It is significant that the dissipative kernels have negative
contributions and are possible to suppress the glassy elastic term.
The intuitive picture of this effect will be discussed below.

Accordingly, the second projection operator
$\mathcal{P}_{\mathrm{mc}}(t)$ is applied to the time correlation
$\langle A(s)\Omega(0) \rangle$ in Eq.~(\ref{eq:ss}), together
with the factorization approximation.
Then we obtain a quadratic form of the time correlation functions.
Again, the explicit forms of the steady-state formulas for $\sigma_{xy}$
and $\mathcal{R}$ depend on $\rho_{\mathrm{ini}}(\bm{\Gamma})$, but
exhibit the following forms in general,
\vspace{-1.0em}
\begin{eqnarray}
(\sigma_{xy})_{\mathrm{SS}} 
=
(\sigma_{xy}^{(\mathrm{el})})_{\mathrm{SS}} 
+
\frac{\zeta}{m\dot\gamma}
(\sigma_{xy}^{(\mathrm{vis})})_{\mathrm{SS}},
\hspace{1em}
\mathcal{R}_{\mathrm{SS}}
=
\frac{\zeta}{m\dot\gamma}
\left[ 
\mathcal{R}^{(\mathrm{loc})}_{\mathrm{SS}}
+
\Delta\mathcal{R}_{\mathrm{SS}}
\right].
\label{eq:sxyR}
\vspace{-0.5em}
\end{eqnarray}
Here, $(\sigma_{xy}^{(\mathrm{el})})_{\mathrm{SS}}$ and
$\zeta(\sigma_{xy}^{(\mathrm{vis})})_{\mathrm{SS}}/(m\dot\gamma)$ are
the elastic and the dissipative components of the shear stress, and
$\mathcal{R}^{(\mathrm{loc})}_{\mathrm{SS}}$ and $\Delta
\mathcal{R}_{\mathrm{SS}}$ are the local and the non-local
contributions, which correspond to the first and the second terms of
Eq.~(\ref{eq:ss}), respectively.

\vspace{-1.0em}
\subsection{Canonical Initial Distribution}

To proceed to explicit calculations, it is necessary to specify the
initial distribution function, $\rho_{\mathrm{ini}}(\bm{\Gamma})$.
We consider the following form,
$%\begin{eqnarray}
\rho_{\mathrm{ini}}(\bm{\Gamma})
=
%\frac{e^{-I(\bm{\Gamma})}}{\int d\bm{\Gamma} e^{-I(\bm{\Gamma})}},
e^{-I(\bm{\Gamma})} / \int d\bm{\Gamma} e^{-I(\bm{\Gamma})},
%\label{eq:rho_ini_general}
$ %\end{eqnarray}
where $I(\bm{\Gamma})$ is the effective potential.
Then, from Eq.~(\ref{eq:Omega_def}), the work function
$\Omega(\bm{\Gamma})$ is expressed as
\begin{eqnarray}
\Omega(\bm{\Gamma})
=
i\mathcal{L}(\bm{\Gamma})I(\bm{\Gamma})
-
\Lambda(\bm{\Gamma})
=
\dot{I}(\bm{\Gamma})
-
\Lambda(\bm{\Gamma}).
\label{eq:Omega_general}
\end{eqnarray}

The most naive choice of $I(\bm{\Gamma})$ is 
\vspace{-0.5em}
\begin{eqnarray}
I(\bm{\Gamma}) 
=
\frac{H_{0}(\bm{\Gamma})}{T},
\label{eq:I_can}
\end{eqnarray}
which corresponds to the canonical initial distribution.
In this case, the system is equilibrated at temperature $T$ at
$t<t_{0}=0$, and shear and dissipation are switched on at $t=t_{0}=0$.
In this section, we show the results for the case of canonical initial
distribution and discuss its validity. 

\vspace{-0.5em}
\subsubsection{Interpretation and Determination of the Temperature}

A simple feature of $\rho_{\mathrm{ini}}(\bm{\Gamma})$ under
$I(\bm{\Gamma})$ of Eq.~(\ref{eq:I_can}) is that the momentum integral
in Eq.~(\ref{eq:ave_def}) is reduced to a Gaussian which can be
performed.
Then, the canonical temperature $T$ in Eq.~(\ref{eq:I_can}) appears in
the coefficients of the MCT equations and the memory kernels.
This poses a problem, since the dynamics governed by the MCT equations,
and hence the steady-state averages, apparently depend on $T$, while it
is expected to be independent from physical grounds.
The origin of this problem resides in our treatment where the dynamics
is projected onto the density and the current density fluctuations, but
not on the temperature fluctuation, and hence the time evolution of the
temperature is not included.

To remedy this problem, we identify $T$ with the steady-state
temperature, and determine it from the energy balance condition,
Eq.~(\ref{eq:ss2}).
As shown below, $T$ appears in the coefficients of the time correlation
functions via the projection operators; 
that is, it appears in MCT equations, Eqs.~(\ref{eq:Phi}) and
(\ref{eq:Psi}), as well as in the energy balance condition,
Eq.~(\ref{eq:ss2}).
We solve MCT equations iteratively until the temperature $T$ is
self-consistently determined by the energy balance condition.
The initial conditions for Eqs.~(\ref{eq:Phi}) and (\ref{eq:Psi}) are
$\Phi_{q}(0)=S_{q}$, $\frac{d}{dt} \Phi_{q}(0)=0$ and $\Psi_{q}(0)=0$,
$\frac{d}{dt}\Psi_{q}(0)=v_T^2$.

\vspace{-0.5em}
\subsubsection{Explicit Expressions}

For the case of the canonical initial distribution, we can explicitly
calculate the coefficients $Z_{q}$, $A_{q}$, and
$A_{q}^{\lambda\lambda}$ in Eqs.~(\ref{eq:Phi}), (\ref{eq:Psi}), and the
coefficients which appear in the memory kernels $M_{\bm{q}}(t)$,
$M_{\bm{q}}^{\lambda\lambda}(t)$ as well as in the steady-state formulas
for $\sigma_{xy}$ and $\mathcal{R}$, i.e. Eqs.~(\ref{eq:sxyR}).

{\it Coefficients--}
The coefficients $Z_{q}$, $A_{q}$, and $A_{q}^{\lambda\lambda}$ are
given by
$%\begin{eqnarray}
Z_{q}
=
%\frac{v_T^2 q^2}{S_{q}},
v_T^2 q^2/S_{q},
$
$A_{q}
=
\frac{4\pi}{3}
%\frac{\zeta_{H}}{m}
(\zeta_{H}/m)
\left[ 1 - j_{0}(qd) + 2j_{2}(qd)\right],
$
and
$A_{q}^{\lambda\lambda}
=
4\pi
%\frac{\zeta_{H}}{m}
(\zeta_{H}/m)
\left[ 1 - j_{0}(qd) \right],
$
where $v_{T} = \sqrt{T/m}$ is the thermal velocity, $j_{l}(qd)$ is the
spherical Bessel function of the $l$-th order, and
\vspace{-0.5em}
\begin{eqnarray}
\frac{\zeta_{H}}{m}
=
4\sqrt{\pi}
\left( 1-e^2 \right)
g(d) nd^2
\sqrt{\frac{T}{m}}
\end{eqnarray} 
is the effective frequency of the dissipation in the hard-core limit.
Here, $e$ is the normal restitution coefficient, and $g(d)$ is the
contact value of the equilibrium radial distribution function.
We take this limit because we address the rheology below the jamming
transition density.

{\it Memory Kernels--}
The memory kernel for $\Phi_{q}(t)$, i.e. $M_{\bm{q}}(t)$, is given up
to linear order in $\zeta_{H}/(m\dot\gamma)$ by
$%\begin{equation}
M_{\bm{q}}(t)
=
M_{\bm{q}}^{(\mathrm{el})}(t)
+
%\frac{\zeta}{m\dot\gamma}
(\zeta_{H}/m\dot\gamma)
\sum_{i=1}^{2}
M_{\bm{q}}^{(\mathrm{vis}i)}(t),
$ %\end{equation}
where $M_{\bm{q}}^{(\mathrm{el})}(t)$,
$M_{\bm{q}}^{(\mathrm{vis}1)}(t)$, and $M_{\bm{q}}^{(\mathrm{vis}2)}(t)$
are given by
\begin{eqnarray}
&&
\hspace{-2em}
M_{\bm{q}}^{(\mathrm{el})}(t)
=
\frac{nv_{T}^2}{32\pi^2 q^3}
\int_{0}^{\infty} dk \, k
\int_{|q-k|}^{q+k} dp \, p
%\nonumber \\
%%
%&&
%\times
V_{M}^{(\mathrm{el})}(\bar{k}(t),\bar{p}(t))
V_{M}^{(\mathrm{el})}(k,p)
\Phi_{k}(t) \Phi_{p}(t),
\\
&&
\hspace{-2em}
M_{\bm{q}}^{(\mathrm{vis}1)}(t)
=
\frac{\dot\gamma}{4\pi mq^2}
\int_{0}^{\infty} dk \, k
\int_{|q-k|}^{q+k} dp
%\nonumber \\
%%
%&&
%\times
V_{M}^{(\mathrm{vis})}(\bar{k}(t),\bar{p}(t),\alpha(t),\beta(t))
V_{M}^{(\mathrm{el})}(k,p)
%\nonumber \\
%%
%&&
%\times
\Phi_{k}(t) 
%\frac{d}{dt}\Phi_{p}(t),
\dot{\Phi}_{p}(t),
\\
&&
\hspace{-2em}
M_{\bm{q}}^{(\mathrm{vis}2)}(t)
=
-\frac{\dot\gamma}{4\pi mq^2}
\int_{0}^{\infty} dk \, k
\int_{|q-k|}^{q+k} dp \, p^2
%\nonumber \\
%%
%&&
%\times
V_{M}^{(\mathrm{el})}(\bar{k}(t),\bar{p}(t))
V_{M}^{(\mathrm{vis})}(k,p,\alpha,\beta)
%\nonumber \\
%%
%&&
%\times
\Phi_{k}(t) 
\Psi_{p}(t).
\hspace{1em}
\end{eqnarray}
Here, $\bar{k}(t) = kh(\dot\gamma t)$ with $h(x) = \sqrt{1 + x^2/3}$ is
the modulus of the advected wavevector in the isotropic approximation,
and the vertex functions are given by
\begin{eqnarray}
%&&
%\hspace{-2em}
&
V_{M}^{(\mathrm{el})}(k,p)
=
c_{k}\left( q^2 + k^2 - p^2\right) 
+
c_{p}\left( q^2 + p^2 - k^2\right),
&
\\
%
%&&
&
\hspace{-1em}
V_{M}^{(\mathrm{vis})}(k,p,\alpha,\beta)
=
\frac{1}{2} \sin\alpha \sin(\alpha+\beta)
\cdot mI^{(1)}(k)
%\nonumber \\
%%
%&&
%\times
\left[
\cos\alpha \cos(\alpha+\beta)
-
\frac{1}{2} \sin\alpha \sin(\alpha+\beta)
\right]
\cdot mI^{(2)}(k)
%\nonumber \\
%%
%&&
-
\cos\beta 
\cdot mI^{(2)}(p),
&
\hspace{2em}
\end{eqnarray}
where $I^{(1)}(q)=2\left[ 1 - j_{0}(qd)\right]$ and
$I^{(2)}(q)=\frac{2}{3}\left[ 1 - j_{0}(qd) + 2 j_{2}(qd)\right]$, and
the angles $\alpha$, $\beta$ are defined by
$\cos\alpha=(q^2+k^2-p^2)/(2qk)$, $\cos\beta=(q^2+p^2-k^2)/(2qp)$.
Note that $V_M^{\mathrm{(el)}}(k,p)$ is the vertex function of the
equilibrium MCT \cite{Goetze}.
Note also that $\Psi_{q}(t)$ couples to the MCT equation of
$\Phi_{q}(t)$, Eq.~(\ref{eq:Phi}), only via
$M_{\bm{q}}^{(\mathrm{vis}2)}(t)$, so $\Psi_{q}(t)$ decouples from
$\Phi_{q}(t)$ in the limit $\zeta\to 0$.
Hence, our theory reduces to the sheared MCT of thermal glassy systems
in the limit $\zeta \to 0$, together with appropriate thermostats,
e.g. Gaussian Isokinetic thermostat, which controls the temperature.

Similarly, the memory kernel for $\Psi_{q}(t)$,
i.e. $M_{\bm{q}}^{\lambda\lambda}(t)$, is given up to linear order in
$\zeta_{H}/(m\dot\gamma)$ by
$%\begin{equation}
M_{\bm{q}}^{\lambda\lambda}(t) 
=
M_{\bm{q}}^{(\mathrm{el})\lambda\lambda}(t) 
+
%\frac{\zeta_{H}}{m \dot\gamma}
(\zeta_{H}/m\dot\gamma)
\sum_{i=1}^{2}
M_{\bm{q}}^{(\mathrm{vis}i)\lambda\lambda}(t),
$ %\end{equation}
where $M_{\bm{q}}^{(\mathrm{el})\lambda\lambda}(t)$,
$M_{\bm{q}}^{(\mathrm{vis}1)\lambda\lambda}(t)$, and
$M_{\bm{q}}^{(\mathrm{vis}2)\lambda\lambda}(t)$ are 
given by
\begin{eqnarray}
&&
\hspace{-1.5em}
M_{\bm{q}}^{(\mathrm{el})\lambda\lambda}(t) 
=
\frac{n v_T^2}{8\pi^2 q} 
\int_{0}^{\infty} dk \, k
\int_{\left| q-k \right|}^{q+k} dp \, p
%\nonumber \\
%%
%&&
%\times
W^{(\mathrm{el})}(\bar{k}(t), \bar{p}(t), k,p,\alpha,\beta)
\Phi_{k}(t) \Phi_{p}(t),
\\
%\end{eqnarray}
%%
%\begin{eqnarray}
&&
\hspace{-1.5em}
M_{\bm{q}}^{(\mathrm{vis}1)\lambda\lambda}(t) 
=
\frac{\dot\gamma}{2\pi m q}
\int_{0}^{\infty} dk \, k
\int_{\left| q-k \right|}^{q+k} dp 
%\nonumber \\
%%
%&&
%\times
\left[
k c_{k}
W^{(\mathrm{vis}1)}(\bar{k}(t),\bar{p}(t), \alpha(t), \beta(t))
%\right.
%\nonumber \\
%%
%&&
%\left.
+
p c_{p}
W^{(\mathrm{vis}2)}(\bar{k}(t),\bar{p}(t), \alpha(t), \beta(t))
\right]
%\nonumber \\
%%
%&&
%\times
\Phi_{k}(t) 
\dot{\Phi}_{p}(t),
\hspace{1.5em}
\\
%\nonumber \\ \\
%\end{eqnarray}
%% 
%\begin{eqnarray}
&&
\hspace{-1.5em}
M_{\bm{q}}^{(\mathrm{vis}2)\lambda\lambda}(t) 
=
-\frac{\dot\gamma}{2\pi m q}
\int_{0}^{\infty} dk \, k
\int_{\left| q-k \right|}^{q+k} dp \, p^2
%\nonumber \\
%%
%&&
%\times
\left\{
k c_{\bar{k}(t)}
W^{(\mathrm{vis}1)}(k,p,\alpha,\beta)
+
p c_{\bar{p}(t)}
%\right.
%\nonumber \\
%%
%&&
%\left.
\left[
W^{(\mathrm{vis}2)}(k,p,\alpha,\beta)
+
\frac{1}{2}
(\dot\gamma t)^2
W^{(\mathrm{vis}3)}(k,\alpha,\beta)
\right]
\right\}
\nonumber \\
&&
\times
\Phi_{k}(t) \Psi_{p}(t).
\end{eqnarray}
The detailed expressions for the coefficients $W^{(\mathrm{el})}$,
$W^{(\mathrm{vis}1)}$, $W^{(\mathrm{vis}2)}$, and $W^{(\mathrm{vis}3)}$
are omitted here.
%
%The coefficients $W^{(\mathrm{el})}$, $W^{(\mathrm{vis}1)}$,
%$W^{(\mathrm{vis}2)}$, and $W^{(\mathrm{vis}3)}$ are given by
%%
%\begin{eqnarray}
%&&
%\hspace{-1em}
%W^{(\mathrm{el})}(\bar{k}(t),\bar{p}(t),k,p,\alpha,\beta) 
%=
%k^2 c_{k} c_{\bar{k}(t)} 
%+
%kp \cos (\alpha+\beta)
%%\nonumber \\
%%%
%%&&
%%\times
%\left( c_{k} c_{\bar{p}(t)} + c_{\bar{k}(t)} c_{p} \right)
%+
%p^2 c_{p} c_{\bar{p}(t)},
%\\
%%\end{eqnarray}
%%%
%%\begin{eqnarray}
%&&
%\hspace{-1em}
%W^{(\mathrm{vis}1)}(k,p,\alpha,\beta)
%=
%\cos(\alpha+\beta) \left[ m I^{(2)}(k) - mI^{(2)}(p)\right],
%\\
%%\end{eqnarray}
%%%
%%\begin{eqnarray}
%&&
%\hspace{-1em}
%W^{(\mathrm{vis}2)}(k,p,\alpha,\beta) 
%=
%\frac{1}{2} \sin^2(\alpha+\beta)
%\cdot mI^{(1)}(k)
%%\nonumber \\
%%%
%%&&
%+
%\frac{3\cos^2(\alpha+\beta)-1}{2}
%\cdot mI^{(2)}(k)
%- 
%mI^{(2)}(p),
%\\
%%
%&&
%\hspace{-1em}
%W^{(\mathrm{vis}3)}(k,\alpha,\beta)  
%=
%\frac{1}{2} \sin^2(\alpha+\beta)
%\left[ mI^{(1)}(k) - mI^{(2)}(k) \right].
%\end{eqnarray}
%%\end{widetext}
%%
Note that $M_{\bm{q}}^{\lambda\lambda}(t)$ is the trace of the tensor
$M_{\bm{q}}^{\mu\nu}(t)$, and hence its coefficients cannot be expressed
as a product of two vertex functions, as is the case for $M_{\bm{q}}(t)$.

{\it Work Function and the Steady-State Formula--}
The explicit form of the work function can also be calculated from
Eqs.~(\ref{eq:Omega_general}) and (\ref{eq:I_can}) as
$%\begin{eqnarray}
\Omega(\bm{\Gamma}) 
=
-\beta \dot\gamma V \sigma_{xy}(\bm{\Gamma})
-
2\beta \mathcal{R}(\bm{\Gamma})
-
\Lambda(\bm{\Gamma}),
%\label{eq:Omega_eq}
$ %\end{eqnarray}
where $\beta = 1/T$ is the inverse temperature.
This expression can be identically rewritten as
\begin{equation}
\Omega(\bm{\Gamma}) 
=
-
\beta \dot\gamma V \sigma_{xy}^{(\mathrm{el})}(\bm{\Gamma})
+
\beta \dot\gamma V \sigma_{xy}^{(\mathrm{vis})(1)}(\bm{\Gamma})
-
2\beta \Delta\mathcal{R}(\bm{\Gamma}),
\label{eq:Omega_eq2}
\end{equation}
where $\sigma_{xy}^{(\mathrm{el})}(\bm{\Gamma})$,
$\sigma_{xy}^{(\mathrm{vis})(1)}(\bm{\Gamma})$, and
$\Delta\mathcal{R}(\bm{\Gamma})$ are given by
%
%\begin{eqnarray}
%&&
$\sigma_{xy}^{(\mathrm{el})}(\bm{\Gamma}) 
=
\frac{1}{V}
\sum_{i=1}^{N}
\left[
%\frac{p_{i}^x p_{i}^y}{m}
p_{i}^x p_{i}^y/m
+
y_{i} F_{i}^{(\mathrm{el})\,x}
\right],
$
%
%&&
$\sigma_{xy}^{(\mathrm{vis})(1)}(\bm{\Gamma})
=
-\frac{\zeta_{H}}{2V}
\sum_{\langle i,j \rangle}
\left( \frac{\bm{p}_{ij}}{m} \cdot \hat{\bm{r}}_{ij}\right)  
\hat{x}_{ij} \hat{y}_{ij} r_{ij}
\Theta(d-r_{ij})
$,
and
\vspace{-1.0em}
\begin{eqnarray}
\Delta\mathcal{R}(\bm{\Gamma})
=
\mathcal{R}^{(1)}(\bm{\Gamma})
+
\frac{T}{2}
\Lambda(\bm{\Gamma}),
\label{eq:DeltaR}
\vspace{-0.5em}
\end{eqnarray}
with
$%\begin{eqnarray}
\mathcal{R}^{(1)}(\bm{\Gamma})
=
\frac{\zeta_{H}}{4}
%(\zeta_{H}/4)
\sum_{\langle i,j \rangle}
\left( \frac{\bm{p}_{ij}}{m} \cdot \hat{\bm{r}}_{ij}\right)^{2}
\Theta(d-r_{ij}).  
$ %\end{eqnarray}
Note that $\Delta\mathcal{R}(\bm{\Gamma})$ is the purely dissipative
contribution which exists in the absense of shear, among which
$\mathcal{R}^{(1)}(\bm{\Gamma})$ is the contribution from the energy
dissipation due to inelastic collisions.

The steady-state shear stress
%, up to linear order in $\zeta/(m\dot\gamma)$, 
reads
$%\begin{equation}
\left( \sigma_{xy} \right)_{\mathrm{SS}}
=
(\sigma_{xy}^{(\mathrm{el})})_{\mathrm{SS}} 
+
%\frac{\zeta_{H}}{m\dot\gamma}
(\zeta_{H}/m\dot\gamma)
\sum_{i=1}^{2}
(\sigma_{xy}^{(\mathrm{vis}i)})_{\mathrm{SS}},
$ %\end{equation}
where $(\sigma_{xy}^{(\mathrm{el})})_{\mathrm{SS}}$,
$(\sigma_{xy}^{(\mathrm{vis}1)})_{\mathrm{SS}}$, and
$(\sigma_{xy}^{(\mathrm{vis}2)})_{\mathrm{SS}}$ are given by
%
%\begin{widetext}
%
\vspace{-0.5em}
\begin{eqnarray}
&&
\hspace{-4em}
(\sigma_{xy}^{(\mathrm{el})})_{\mathrm{SS}} 
=
\frac{\dot\gamma T}{60 \pi^2} 
\int_{0}^{\infty} \frac{dt}{h(\dot\gamma t)}
\int_{0}^{\infty} dk \, k^4 
%\nonumber \\
%%
%&&
%\times
V_{\sigma}^{(\mathrm{el})}(\bar{k}(t)) %\frac{W_{\bar{k}(t)}}{S_{\bar{k}(t)}}
V_{\sigma}^{(\mathrm{el})}(k) %\frac{W_{k}}{S_{k}}
\Phi_{k}(t)^2,
\label{eq:sxy_el}
\\
%\end{eqnarray}
%%
%\begin{eqnarray}
&&
\hspace{-4em}
(\sigma_{xy}^{(\mathrm{vis}1)})_{\mathrm{SS}} 
=
- \frac{\dot\gamma^2}{60\pi^2} 
\int_{0}^{\infty} \frac{dt}{h(\dot\gamma t)^3}
\int_{0}^{\infty} dk \, k^2
%\nonumber \\
%%
%&&
%\times
V_{\sigma}^{(\mathrm{vis})}(\bar{k}(t)) 
V_{\sigma}^{(\mathrm{el})}(k) % \frac{W_{k}}{S_{k}}
\Phi_{k}(t)
%\frac{d}{dt} \Phi_{k}(t),
\dot{\Phi}_{k}(t),
\\
\label{eq:sxy_vis1}
%\end{eqnarray}
%%
%\begin{eqnarray}
&&
\hspace{-4em}
(\sigma_{xy}^{(\mathrm{vis}2)})_{\mathrm{SS}}  
=
\frac{\dot\gamma^2}{60\pi^2}
\int_{0}^{\infty} \frac{dt}{h(\dot\gamma t)}
\int_{0}^{\infty} dk \, k^4
%\nonumber \\
%%
%&&
%\times
V_{\sigma}^{(\mathrm{el})}(\bar{k}(t)) %\frac{W_{\bar{k}(t)}}{S_{\bar{k}(t)}}
V_{\sigma}^{(\mathrm{vis})}(k)
\Phi_{k}(t)
\Psi_{k}(t).
\label{eq:sxy_vis2}
\end{eqnarray}
%
%\end{widetext}
%
The vertex functions $V_{\sigma}^{(\mathrm{el})}(q)$ and
$V_{\sigma}^{(\mathrm{vis})}(q)$ are given by
$%\begin{eqnarray}
V_{\sigma}^{(\mathrm{el})}(q)
=
%\frac{1}{S_{q}}
%\frac{\partial \ln S_{q}}{\partial q}
(\partial \ln S_{q}/\partial q)/S_{q}
$
and
$V_{\sigma}^{(\mathrm{vis})}(q)
=
-\frac{1}{2} m 
\left[I_{\sigma}^{(1)}(q) -3 I_{\sigma}^{(2)}(q)
\right],
$
with
$
I_{\sigma}^{(1)}(q)
\equiv
2 d j_1(qd)
$
and
$I_{\sigma}^{(2)}(q)
\equiv
\frac{4}{5} d
\left[ \frac{3}{2} j_1(qd) - j_3(qd) \right].
$
Note that $(\sigma_{xy}^{(\mathrm{el})})_{\mathrm{SS}}$,
Eq.~(\ref{eq:sxy_el}), is coincident to the steady-state shear stress in
the sheared MCT of thermal glassy systems~\cite{FC2002, MRY2004}.

The steady-state energy dissipation rate
%, up to linear order in $\zeta/(m\dot\gamma)$, 
is given by
\vspace{-0.5em}
\begin{equation}
\mathcal{R}_{\mathrm{SS}}
=
\frac{\zeta_{H}}{m\dot\gamma}
\left[
\mathcal{R}^{(\mathrm{loc})}_{\mathrm{SS}}
+
\sum_{i=1}^{3}
\mathcal{R}_{\mathrm{SS}}^{(i)}
\right],
\label{eq:Rss}
\end{equation}
where $\mathcal{R}^{(\mathrm{loc})}_{\mathrm{SS}}$ is the local
contribution,
\vspace{-0.5em}
\begin{eqnarray}
\mathcal{R}^{(\mathrm{loc})}_{\mathrm{SS}} 
&=&
\frac{1}{2} \dot\gamma g(d) N n d^3 T,
\label{Eq:Req}
\end{eqnarray}
%\vspace{-0.5em}
%
and $\mathcal{R}^{(i)}_{\mathrm{SS}}$ ($i=1,2,3$) are the non-local
contributions, 
%
%
%\begin{widetext}
\begin{eqnarray}
&&
\hspace{-3em}
\mathcal{R}^{(1)}_{\mathrm{SS}} 
=
\frac{\dot\gamma^2 T}{6\pi^2} V
\int_{0}^{\infty} dt
\frac{\dot\gamma t}{h(\dot\gamma t)}
\int_{0}^{\infty} dk \, k^4
%\nonumber \\
%%
%&&
%\times
\frac{1}{2} I_{\sigma}^{(1)}(k)
V_{\sigma}^{(\mathrm{el})}(\bar{k}(t)) %\frac{W_{\bar{k}(t)}}{S_{\bar{k}(t)}}
\Phi_{\bar{k}(t)}(t)^2,
\label{Eq:R1}
\\
%\end{eqnarray}
%%
%\begin{eqnarray}
&&
\hspace{-3em}
\mathcal{R}^{(2)}_{\mathrm{SS}} 
=
- \frac{\dot\gamma^3}{30\pi^2} V
\int_{0}^{\infty} 
\frac{dt}{h(\dot\gamma t)^3}
\int_{0}^{\infty} dk \, k^2
%\nonumber \\
%%
%&&
%\times
V_{\sigma}^{(\mathrm{vis})}(\bar{k}(t))
V_{\sigma}^{(\mathrm{el})}(k) %\frac{W_{k}}{S_{k}}
\Phi_{k}(t)
%\frac{d}{dt} \Phi_{k}(t),
\dot{\Phi}_{k}(t),
\label{Eq:R2}
\\
%\end{eqnarray}
%%
%\begin{eqnarray}
&&
\hspace{-3em}
\mathcal{R}^{(3)}_{\mathrm{SS}}
=
\frac{\dot\gamma^3}{3\pi^2} V
\int_{0}^{\infty} dt
\frac{\dot\gamma t}{h(\dot\gamma t)}
\int_{0}^{\infty} dk \, k^4
%\nonumber \\
%%
%&&
%\times
\frac{1}{2} I_{\sigma}^{(1)}(k)
V_{\sigma}^{(\mathrm{vis})}(\bar{k}(t))
\Phi_{\bar{k}(t)}(t)
\Psi_{\bar{k}(t)}(t).
\label{Eq:R3}
\end{eqnarray}
%\end{widetext}
%
%
%The vertex functions $V_{\sigma}^{(\mathrm{el})}(q)$,
%$V_{\sigma}^{(\mathrm{vis})}(q)$, and $I_{\sigma}^{(1)}(q)$ are given by
%Eqs.~(\ref{Eq:W}), (\ref{Eq:Vvis5}), and (\ref{Eq:Is1}).
%
%A shift in the integration variable $k \to k(t) \simeq \bar{k}(t)$ is
%performed in Eqs.~(\ref{Eq:R1}) and (\ref{Eq:R3}) to extract the
%anisotropic terms to perform the isotropic approximation.
%
Note that $\mathcal{R}^{(\mathrm{loc})}_{\mathrm{SS}}$,
Eq.~(\ref{Eq:Req}), is formally coinicident to the kinetic theory
expression of the energy dissipation rate \cite{GD1999}.
%
%Hence, the remaining three terms $\mathcal{R}^{(i)}$ ($i=1,2,3$)
%correspond to the non-canonical contribution, which is not included in
%the kinetic theory and hence is specific to our work.
%
%However, the physical contents are quite different.
%, in parallel to the discussion in section \ref{sec:hcl}.
%

\vspace{-0.5em}
\subsubsection{Numerical Calculations}

%In our calculation of MCT, the unit of mass, length, and time are chosen
%as $m$, $d$, and $\dot\gamma^{-1}$. 
%
The inputs necessary for the calculation are $S_{q}$ and $g(d)$, both at
equilibrium.
We adopt for $S_{q}$ the Percus-Yevick solution for hard spheres
\cite{HM} ($\varphi \leq 0.52$) and the numerical results from the MD
simulation ($\varphi > 0.52$), and for $g(d)$ the interpolation formula
valid in the range $0.49 < \varphi < 0.64$ \cite{Torquato1995}.

In order to verify the validity of our theory, we also perform MD
simulations of the Sllod equations for soft spheres, Eqs.~(\ref{eq:r})
and (\ref{eq:p}), under the Lees-Edwards boundary condition \cite{EM}.
The equations are integrated by the Verlet algorithm.
The conditions are: the time step $\Delta t = 0.01 \sqrt{m/\kappa}$, where
$\kappa$ is the spring constant, the number of particles $N=2000$, and the
dimensionless shear rate $\dot\gamma \sqrt{m/\kappa} = 1.0\times 10^{-4}$.
The values of $\zeta$ are determined by $\kappa$ and $e$.
The results are averaged from 10 independent samples.

\begin{figure*}[tbh]
%\begin{center}
\includegraphics[width=8.5cm]{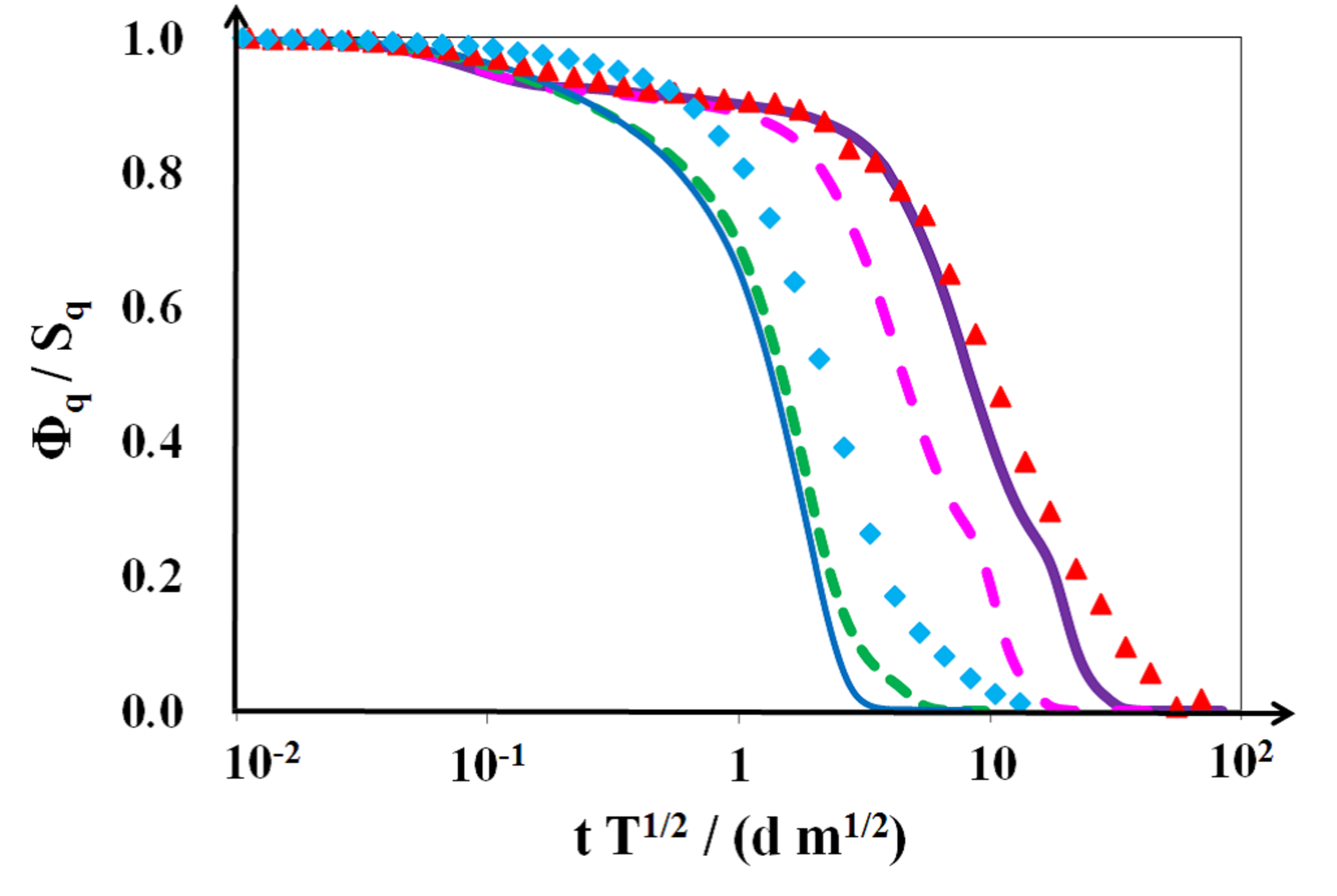} 
\includegraphics[width=8.5cm]{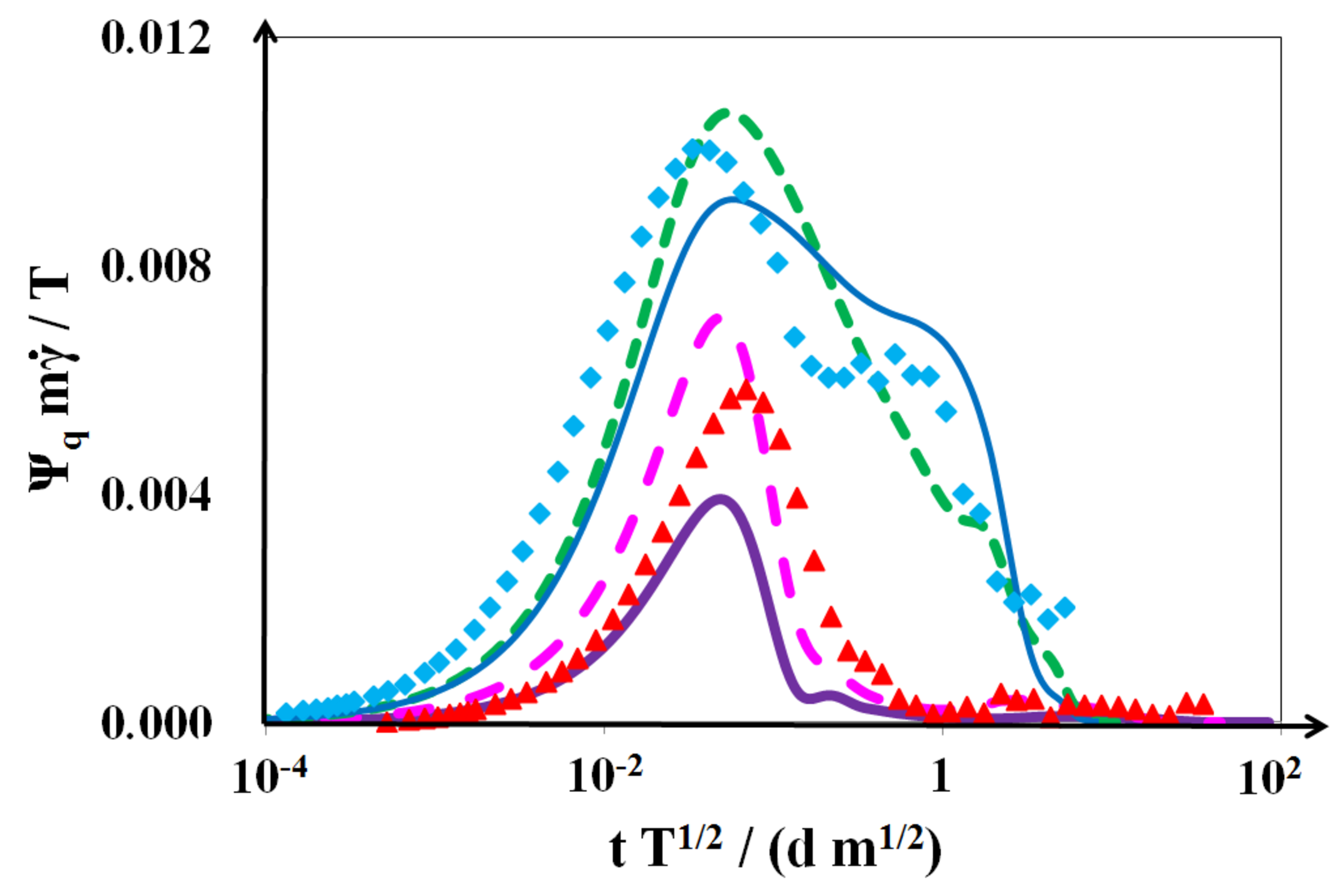} 
%\end{center} 
\vspace{-2.0em}
\caption {The relaxation of the time correlation
functions. The left (right) panel is for the density-density
(density-current) time correlation function $\Phi_{q}(t)
(\Psi_{q}(t))$. The results of MCT for $\varphi = 0.52$ and $e$ =0.99,
0.98, 0.94, 0.92 are shown in thick solid, broken,
dashed, and thin solid lines.  The results of MD for
$\varphi = 0.63$ and $e$ =0.9999, 0.999 are shown in triangles and
diamonds.  The wave number $qd = 7.4$ is chosen for both MCT and
MD, which corresponds to the first peak of the static structure factor
measured in MD. The results of MD for the density-current correlation
$\Psi_{q}(t)$ is multiplied by a factor 30 to be comparable with the
results of MCT in a single figure.}
\label{Fig:TCF}
\vspace{-1em}
\end{figure*}

{\it Time correlation functions--}
The left (right) panel of Fig.~\ref{Fig:TCF} is the relaxation of the
density-density (density-current) time correlation function
$\Phi_{q}(t)$ ($\Psi_{q}(t)$).
The reader is referred to the caption for the conditions.
The result of MCT for $\Phi_{q}(t)$ shows a clear two-step relaxation in
the nearly elastic cases ($e$ = 0.99, 0.98), while it cannot be seen for
the dissipative cases ($e$ = 0.94, 0.92).
This is qualitatively consistent with the result of the MD simulations,
though the plateau is suppressed even for $e=0.999$ in MD.
Note that our results for MCT and MD which exhibit no plateau in
$\Phi_{q}(t)$ are consistent with the results of MD reported previously
\cite{CC2009}.
%\cite{DMB2005, Kumaran2009, CC2009}.
%
The result of MCT for $\Psi_{q}(t)$ shows a clear peak and a shoulder
for the dissipative cases, while it has only a peak in short time scales
for the nearly elastic cases.
This is again qualitatively consistent with MD, though the amplitude of
MD is much smaller than that of MCT.
It is interesting to note that a plateau or a shoulder appears
complementarily in $\Phi_{q}(t)$ and $\Psi_{q}(t)$.
This is an evidence of the significant role of $\Psi_{q}(t)$ in sheared
granular liquids.
From this result, we can derive the following intuitive picture when the
dissipation is significant.
The spheres lose density correlations because they cease to collide with
the surrounding caging spheres at the time scale $\sim m/\zeta_{H}$ due
to inelasticity.
%%%
%
On the other hand, the spheres "memorize" the information of the current
at the onset of shearing and inelastic collisions, since they lose the
kinetic energy more rapidly when they interact with currents which have
larger relative velocity.
The spheres become crowded where the dissipation rate is large, which
results in a density fluctuation; i.e., the density fluctuation is
correlated with the initial current, even at the time scales of slow
motions.
This correlation is eventually destroyed by shearing at the time scale
$\sim \dot\gamma^{-1} > m/\zeta_{H}$.
In contrast, for elastic or driven cases, the information of the initial
current is lost and no density fluctuation is generated.
%%%
%
%On the other hand, the density fluctuation remains correlated with the
%initial current fluctuation, until the cage is destructed by the shear
%at the time scale $\sim \dot\gamma^{-1} > m/\zeta_{H}$.
%

\begin{figure}[tbh]
%\begin{center}
\includegraphics[width=8.5cm]{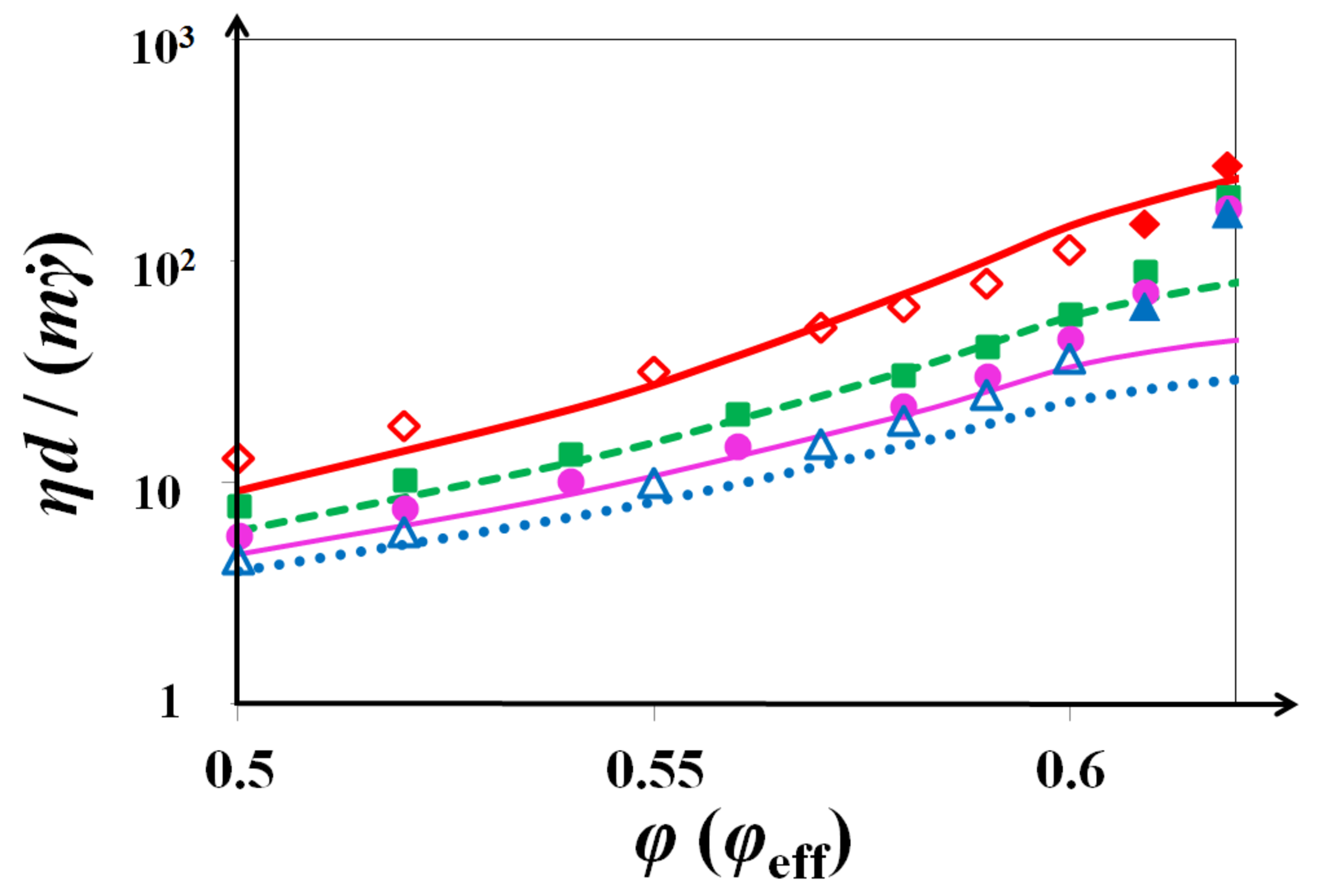} 
\includegraphics[width=8.5cm]{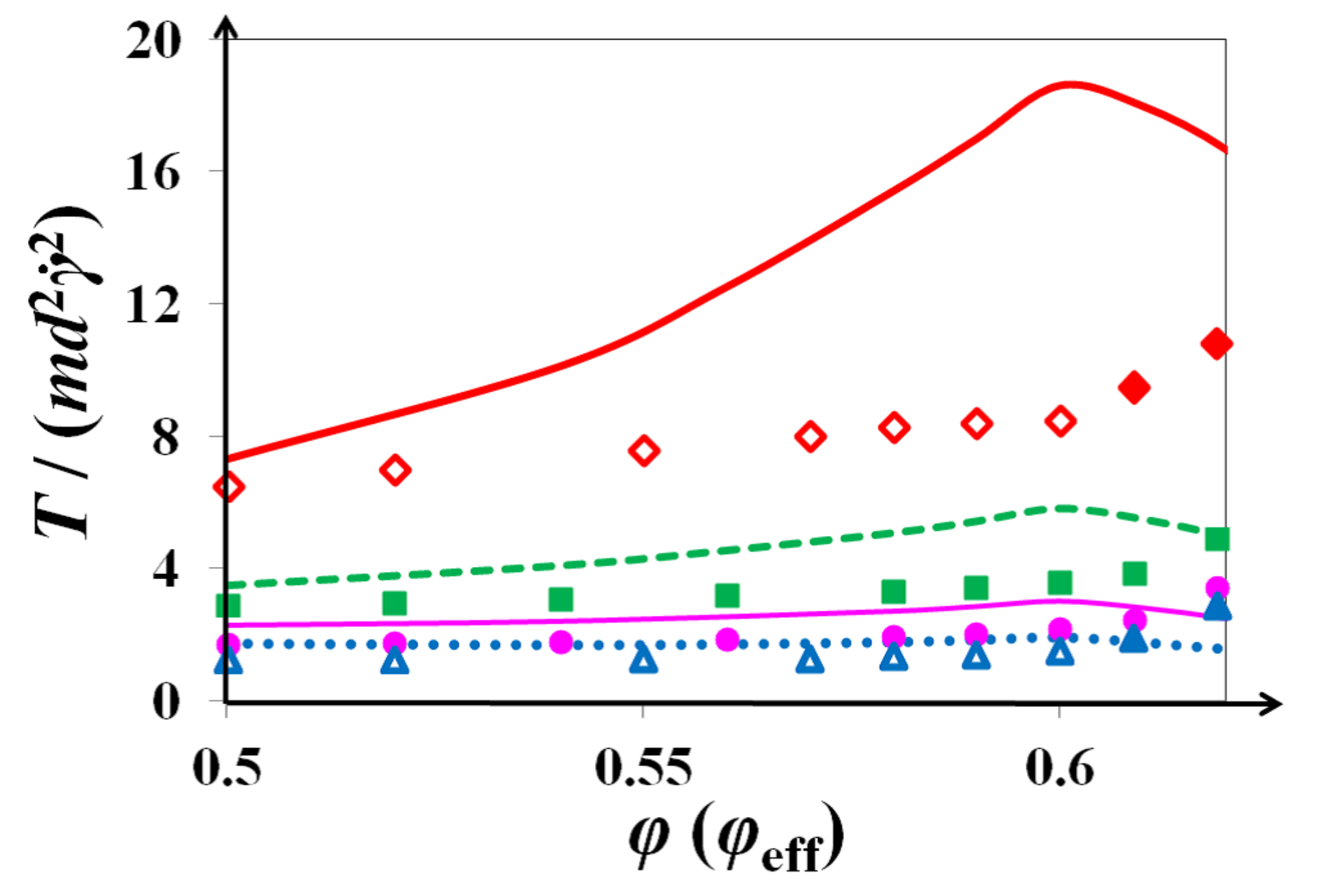} 
%\end{center} 
\vspace{-2em}
\caption{The density dependence of the shear viscosity (left panel) and the
temperature (right panel). 
The results of
MCT for $e$ =0.98, 0.96, 0.94, 0.92 are shown in thick solid, dashed,
thin solid, and dotted lines, where the density is shifted as
$\varphi \to \varphi_{\mathrm{eff}} = \varphi + \Delta\varphi$ with
$\Delta\varphi=0.11$. The results of MD for $e$ =0.98, 0.96, 0.94, 0.92
are shown in diamonds, squares, circles, and triangles. The filled
symbols are from the MD we have performed, and the open symbols are
cited from Ref.~\cite{MN2007}.}
\label{Fig:Viscosity}
\vspace{-1em}
\end{figure}

{\it Shear viscosity--}
The density dependence of the shear viscosity $\eta =
\sigma_{xy}/\dot\gamma$ is shown in the left panel of
Fig.~\ref{Fig:Viscosity}.
In this figure, we also plot the result of MD for hard spheres
\cite{MN2007}.
We find a remarkable agreement between MCT and MD for $\varphi < 0.60$
by using the shift of the density in MCT as $\varphi \to
\varphi_{\mathrm{eff}} \equiv \varphi + \Delta\varphi$ with
$\Delta\varphi = 0.11$.
This leads to the MCT transition at the effective density,
$\varphi_{\mathrm{MCT,eff}} = \varphi_{\mathrm{MCT}} + \Delta\varphi
\simeq 0.626$.
%
%%%
%
%
%%%
%
Although this value is higher than the glass transition density,
$\varphi_{g} \simeq 0.58-0.60$, it is close to 0.62, which is used to
explain the result of the simulations for colloidal hard-sphere
dispersions by MCT, for both the long-time self-diffusion coefficient
\cite{BBN1999} and the shear viscosity \cite{CZC2002}.
The results of MCT approximately scales as $\eta \propto \left(
\varphi_{\mathrm{J}} - \varphi_{\mathrm{eff}} \right)^{-\alpha}$ for
$\varphi_{\mathrm{eff}} < 0.60$, where a cross-over from $\alpha = 3$ to
$1$ can be observed as reported in Ref.~\cite{IB2013}, depending on the
choice of $\varphi_{J}$.
However, for $\varphi_{\mathrm{eff}} > 0.60$, $\alpha$ becomes smaller
and MCT fails to reproduce the divergence observed in MD.
In fact, the result of MD shows a cross-over at $\varphi \simeq 0.60$ to
a stronger divergence as $\varphi \to \varphi_{\mathrm{J}}$.
Comparison with the related works on the cross-over of the exponent from
1 to 4 for 2D disks \cite{OHL2010} and understanding this cross-over is
a future task.
The divergent feature for $\varphi \to \varphi_{\mathrm{J}}$ is expected
to originate in the contact network of the spheres, which is not
considered in the present framework.
Hence, it is reasonable to find a discrepancy between MCT and MD in the
vicinity of the jamming transition density.
%
%
%It is also worth noting that the density-current correlation
%$\Psi_{q}(t)$ plays an important role in evaluating, especially, the energy
%dissipation rate $\mathcal{R}_{\mathrm{SS}}$.
%%
%This is another evidence of the significance of $\Psi_{q}(t)$.
%%
%
%

{\it Temperature--}
The density dependence of the temperature is shown in the right panel of
Fig.~\ref{Fig:Viscosity}, together with the granular temperature
measured in MD.
The results of MCT show that the temperature monotonically increases up
to $\varphi \simeq 0.60$ in accordance with MD, although the magnitudes
are systematically larger.
However, qualitative discrepancies are found for $\varphi > 0.60$, where
$T$ decreases.
%
%This indicates that there is difficulty in identifying $T$ as the
%steady-state granular temperature in this framework. 
%
This indicates that the balance of the shearing and the energy
dissipation, which determines $T$, is invalid for $\varphi > 0.60$ in
this framework .
%

%\vspace{-0.5em}
\section{Discussions}
\subsection{Problem of Canonical Initial Distribution}
First we discuss the problem of the canonical initial distribution.
As discussed in the previous section, we can infer that the correct
balance between the shearing and the dissipation is not realized in MCT.
To inspect in detail, let us observe the steady-state energy dissipation
rate $\mathcal{R}_{\mathrm{SS}}$, Eq.~(\ref{eq:Rss}).
While the local contribution $\zeta_{H}
\mathcal{R}_{\mathrm{SS}}^{(\mathrm{loc})}/(m\dot\gamma)$ is independent
of $\dot\gamma$, the non-local contributions with time integrations,
$\zeta_{H} \mathcal{R}_{\mathrm{SS}}^{(i)} /(m\dot\gamma)$ ($i=1,2,3$),
are proportional to $\dot\gamma$ or $\dot\gamma^2$.
This indicates that the purely dissipative contribution is absent in
$\zeta_{H} \mathcal{R}_{\mathrm{SS}}^{(i)} /(m\dot\gamma)$ ($i=1,2,3$),
which is physically unacceptable, since inelastic collisions are
expected to exhibit non-local, as well as local, time correlations.
%
%%%%%%%%%%%%%%%%%%
%
%
%%%%%%%%%%%%%%%%%%

The origin of this problem can be traced back to the choice of the
canonical initial distribution as follows.
In MCT, the time correlation $\langle A(t)\Omega(0) \rangle$ in the
steady-state formula, Eq.~(\ref{eq:ss}), is approximated as
\begin{equation}
\left\langle
A(t) \Omega(0)
\right\rangle
\approx
\hspace{0.1em}
\left\langle
\left[
\mathcal{U}_{0}(t,0)
\mathcal{P}_{\mathrm{mc}}^{0}(t)
A(t) 
\right]
\left[
\mathcal{P}_{\mathrm{mc}}^{0}(0)
\Omega(0)
\right]
\right\rangle,
\end{equation}
where $\mathcal{U}_{0}(t,0)$ is the time evolution operator in the space
orthogonal to the projected space \cite{SH2013-1}, and
$\mathcal{P}_{\mathrm{mc}}^{0}(t)$ is the second projection operator,
Eq.~(\ref{eq:Pmc}), restricted to zero-wavevector.
The specific feature of the canonical initial distribution is that the
purely dissipative contribution in the work function, i.e. $\Delta
\mathcal{R}$ of Eq.~(\ref{eq:DeltaR}), is projected out by
$\mathcal{P}_{\mathrm{mc}}^{0}$, i.e.
\begin{eqnarray}
\mathcal{P}_{\mathrm{mc}}^{0}(0)
\Delta\mathcal{R}(0) 
=
0,
\end{eqnarray}
which results in 
$%\begin{eqnarray}
\left\langle
A(t) \Delta\mathcal{R}(0)
\right\rangle
=
0. 
$ %\end{eqnarray}
This indicates that the purely dissipative effect is missing in the
steady-state formula.
This observation suggests us to adopt a non-canonical initial
distribution.

In general, it is expected that the steady-state averages are
insensitive to the choice of the initial distribution.
In fact, it is shown explicitly that this is true for some special cases
\cite{HCO2010}. 
However, this feature is violated in the approximate formulation of MCT.
This is because MCT is intended to describe long-time dynamics, and
sacrifices the accuracy for the description of the initial relaxation.
One evidence we have already seen is its dependence on the initial
temperature, which appears in the coefficients.
Thus, the choice of the initial distribution is crucial in MCT for
sheared granular liquids.
A basic idea to remedy this problem is to choose the initial
distribution which is much closer to the steady-state distribution.
To obtain an exact steady-state distribution is of course an extremely
difficult task which is out of our scope at present.
However, we might be able to speculate a valid distribution.

\vspace{-1.0em}
\subsection{Relation to Previous Works}
Next we discuss the relation of this work to the previous related works.
In the former MCTs for thermal sheared glassy systems \cite{FC2002} or
randomly driven granular systems \cite{KSZ2013}, only the projection to
the pair-density modes, i.e. $\mathcal{P}_{nn}$ of
Eq.~(\ref{eq:PnnPnj}), has been included.
In fact, we have already shown in Ref.~\cite{SH2013-2} that the effect
of the projection to the density-current modes, i.e. $\mathcal{P}_{nj}$
of Eq.~(\ref{eq:PnnPnj}), is negligible in thermal sheared underdamped
systems \cite{SH2013-1}.
However, we have figured out that including the correlations to the
density-current modes, or, equivalently, considering $\Psi_{q}(t)$ as
well is crucial to take into account the dissipation effect for sheared
granular liquids.
This can be convinced by observing that the dissipative force is
proportional to the relative velocity of spheres, and hence dissipation
is correlated with current fluctuations.
%
%This fact also implies that the unified picture proposed in
%Ref.~\cite{IBS2012} cannot be applied to dissipative granular materials,
%since the density-current correlation, which is not considered in
%Ref.~\cite{IBS2012}, is significant for them.

\vspace{-1.5em}
\subsection{Scaling of the Shear Stress}

As for the scaling of the shear viscosity with the shear rate, we only
obtain the Bagnold scaling $\eta \propto \dot\gamma$.
This seems to originate from the hard-core limit we have adopted.
To reproduce a departure from the Bagnold scaling, which is significant
near the jamming transition point, we should take into account the
soft-core nature in our theory \cite{IB2013}.
More specifically, 
%the soft-core formulation alone might be insufficient, and 
it might be necessary to further incorporate the information of the
contact networks formed between the particles.
However, constructing a theory based on the generalized Green-Kubo
formula is still promising, because it holds even above the jamming
transition point \cite{HO2013}.

\section{Conclusion}

We have succeeded in predicting the time correlations and the shear
viscosity of dense sheared granular liquids by extending MCT, although
the prediction of time correlations requires refinement to improve
quantitative accuracy.
It has been demonstrated that, in contrast to thermal glassy systems,
the density-current correlation plays an essential role.
Its validity for the prediction of the shear viscosity has been verified
for $\varphi < 0.60$ by comparing it with MD, with the aid of the shift
of the density.
This scheme is expected to be further extended to higher densities
$\varphi > 0.60$ by incorporating the soft-core nature of the spheres.
On the other hand, our theory fails to predict the appropriate tendency
of the granular temperature for $\varphi > 0.60$.
We have discussed that this problem might reside in the choice of the
initial distribution.
To remedy this problem, it seems to be crucial to adopt a distribution
which is expected to be close to the steady-state distribution.
%

%%%%%%%%%%%%%%%%%%%%%%%%%%%%%%%%%%%%%%%%%%%%%%%%
%% BACKMATTER
%%%%%%%%%%%%%%%%%%%%%%%%%%%%%%%%%%%%%%%%%%%%%%%%

\vspace{-0.5em}
\begin{theacknowledgments}
The authors are grateful to S.-H. Chong and M. Otsuki for their
collaboration in the initial stage of this project and extensive
discussions.
They are also grateful to M. Sperl, W. T. Kranz, and M. Fuchs
for fruitful discussions, and T. G. Sano and S. Takada for providing us
the prototype of the program for the MD simulation.
This work is partially supported by the Grant-in-Aid of MEXT (Grant
No. 25287098).
The numerical calculations in this work were carried out at the computer
facilities at the Yukawa Institute.
\end{theacknowledgments}



\vspace{-0.5em}
\bibliographystyle{aipproc}   % if natbib is available

\bibliography{Suzuki-Hayakawa}

\IfFileExists{\jobname.bbl}{}
 {\typeout{}
  \typeout{******************************************}
  \typeout{** Please run "bibtex \jobname" to optain}
  \typeout{** the bibliography and then re-run LaTeX}
  \typeout{** twice to fix the references!}
  \typeout{******************************************}
  \typeout{}
 }

\end{document}